\documentclass[longbibliography,groupedaddress,showpacs,showkeys,amssymb,eqsecnum,aps,nofootinbib,superscriptaddress]{revtex4-2}
\usepackage[pdftex]{graphicx}
\usepackage{amsmath} 
\usepackage{epsf}
\usepackage{color}
\usepackage{verbatim}                                                                         
\usepackage[pdftex]{hyperref}
\newcommand{\be}{\begin{equation}}

\newcommand{\ee}{\end{equation}}                                                                         

\begin{document}

\title{Thermal quantum gravity in a general background gauge}
\author{F. T. Brandt}
\email{fbrandt@usp.br}
\affiliation{Instituto de F\'{\i}sica, Universidade de S\~ao Paulo, S\~ao Paulo, SP 05508-090, Brazil}
\author{J. Frenkel}
\email{josiffrenkel@gmail.com}
\affiliation{Instituto de F\'{\i}sica, Universidade de S\~ao Paulo, S\~ao Paulo, SP 05508-090, Brazil}
\author{D. G. C. McKeon}
\email{dgmckeo2@uwo.ca}
\affiliation{
Department of Applied Mathematics, The University of Western Ontario, London, ON N6A 5B7, Canada}
\affiliation{Department of Mathematics and Computer Science, Algoma University,
Sault St.Marie, ON P6A 2G4, Canada}

\author{G. S. S. Sakoda}    
\email{gustavo.sakoda@usp.br}
\affiliation{Instituto de F\'{\i}sica, Universidade de S\~ao Paulo, S\~ao Paulo, SP 05508-090, Brazil}

\date{\today}
\begin{abstract}
  We calculate in a general background gauge, to one-loop order, the leading logarithmic contribution from the graviton self-energy at finite temperature $T$, extending a previous analysis done at $T=0$.
The result, which has a transverse structure, is applied to evaluate the leading quantum correction of the gravitational vacuum polarization to the Newtonian potential. An analytic expression valid at all temperatures is obtained, which generalizes  the result  obtained earlier at $T=0$. One finds that the magnitude of this quantum correction decreases as the temperature rises.
\end{abstract}                                                                       
\maketitle                     

\section{Introduction}
Classical general relativity is a successful theory which provides a very good description of the gravitational interactions that occur at low energies. There have been many attempts to quantize gravity along the lines of other field theories and it was recognized that general relativity is not renormalizable \cite{r5,r7,r8,r11,r12,r14}. The contributions generated by Feynman loop diagrams to all orders require an infinite number of counter-terms to cancel all ultraviolet divergences, which leads to a lack  of predictability of such a theory at high energies. A point of view now well established in many areas of physics is that physical predictions at low energies, that are well verified experimentally, can be made in non-renormalizable theories. The key ingredient of such  predictions is the fact that these must be made within the context of an effective low energy theory, in powers of the energy divided by some characteristic heavy mass. 
Much work has been done to treat general relativity as an effective field theory \cite{Burgess:2003jk,r1,r3,r4}, which upon quantization may lead to predictive quantum corrections at low energies. A special class of low-energy corrections, involving non-local effects, appears to be quite important. The non-locality is manifest by a non-analytic behavior due, for example, to the presence of logarithmic corrections of the form $\log(-k^2)$, where $k$ is some typical momentum transfer. Because these terms become large for small enough $k^2$, they will yield the leading quantum corrections in the limit $k^2\rightarrow 0$. Such terms arise from long distance propagations of massless gravitons. As shown in Refs \cite{r25,r26}, these  effects lead to calculable finite quantum corrections to the classical gravitational potential. (For an alternative treatment see Ref. \cite{r13New})

In this framework, the background field method \cite{r17,r19,r20,r21,r22,r23} has been much employed 
in the computation of quantum corrections in  quantum gravity since this procedure preserves the gauge invariance of the  background field. It has been first shown by Hooft and Veltman \cite{r5} that on mass-shell, pure  gravity is renormalizable to one-loop order. This analysis has been done in a particular  background gauge, obtained by setting the gauge parameter equal to $1$. In a previous work \cite{ra}, we have examined this calculation in a general background gauge and deduced the corresponding  effective Lagrangian. This result was then applied to evaluate, in this class of gauges, the quantum corrections generated by the gravitational vacuum polarization to the Newtonian potential at zero temperature.

A useful extension of this  approach is the calculation of graviton amplitudes at finite temperature $T$. These are of interest in quantum gravity in their own right as well as for their  cosmological applications \cite{ra1,park1}. It has been shown that the $\log(T^2)$ contributions at high temperature have the same Lorentz covariant form as the $\log(- k^2)$ terms at zero temperature \cite{r28,r29,r30}. The purpose of this work is to extend these results to obtain the logarithmic contributions of the graviton self-energy at any temperature. We find  an analytic  expression which smoothly interpolates between the zero and the high temperature limits [Eq. \eqref{eq17}]. We use this form in the static case $k_0 = 0$, to calculate the thermal corrections to the gravitational potential in a general  background gauge. The corresponding result given in Eq. \eqref{eq26} generalizes the one previously obtained at zero temperature.

In Sec. \ref{sec2}  we outline the properties of thermal quantum gravity in a general background gauge. In Sec. \ref{sec3} we evaluate, to one-loop order, the leading logarithmic contributions of the graviton self-energy at finite temperature. As an application, we calculate in Sec. \ref{sec4} the corresponding quantum correction to the classical gravitational potential. We conclude the paper with a brief discussion in Sec. \ref{sec5}. Some details of the computations are given in the Appendix.

\section{Quantization in a general background gauge}\label{sec2}

The theory of quantum gravity is based on the Einstein-Hilbert Lagrangian 
\be\label{eq1}
{\cal L}_g = \sqrt{-g} \frac{2}{\kappa ^2} R,
\ee
where $R$ is the curvature scalar and $\kappa^2 = 32 \pi G$ ($G$ is Newton's constant).
The metric tensor $g_{\mu\nu}$  is  divided into a classical background field $\bar g_{\mu\nu}$
and a quantum field, $h_{\mu\nu}$ so that
\be\label{eq2}
g_{\mu\nu} = \bar g_{\mu\nu} + \kappa h_{\mu\nu},
\ee
where the background field vanishes at infinity, but is arbitrary elsewhere. 
Expanding the Lagrangian \eqref{eq1} in powers of the quantum field,  one obtains for the  quadratic part the contribution \cite{ra}
\begin{eqnarray}\label{eq3}
 {\cal L}^{(2)}_g & = & \sqrt{-\bar g}\Biggl[{1 \over 2} \bar D_{\alpha} h_{\mu \nu} \bar D^{\alpha}
h^{\mu \nu} - {1 \over 2} \bar D_{\alpha} h \bar D^{\alpha} h + \bar D_{\alpha} h
\bar D_{\beta} h^{\alpha \beta}
-\bar D_{\alpha} h_{\mu \beta} \bar D^{\beta} h^{\mu \alpha}
\nonumber \\ 
&+&  \bar{R} \left(
{1 \over 4} h^2 - {1 \over 2} h_{\mu \nu}h^{\mu \nu} \right)
+  \bar{R}^{\mu \nu} \left(2 h^{\alpha} _{~\mu} h_{\nu \alpha}
- h h_{\mu \nu} \right) \Biggr]. 
\end{eqnarray}
where  $h=h^\lambda_\lambda$, $\bar D_\alpha$ is the covariant derivative with respect to the background field and $\bar R_{\mu\nu}$ is the Ricci tensor associated with the background field.

In order to quantize this theory one must fix the gauge of the quantum field in a way that preserves the gauge invariance under the background field transformation
\be\label{eq4}
\delta \bar g_{\mu\nu} = \omega^\gamma \partial_\gamma \bar g_{\mu\nu}
+\bar g_{\mu\gamma} \partial_\nu \omega^\gamma +\bar g_{\nu\gamma} \partial_\mu \omega^\gamma 
= \bar D_\mu \omega_\nu+\bar D_\nu \omega_\mu ,
\ee
This can be accomplished by introducing the gauge-fixing Lagrangian 
\begin{equation}\label{eq5}
{\cal L}_{gf} = \frac{1}{\xi}\sqrt{-\bar g}\left[\left( \bar D^{\nu} h_{\mu \nu} - {1 \over
  2} \bar D_{\mu} h \right) \left( \bar D_{\sigma}
h^{\mu\sigma} - {1 \over 2} \bar D^{\mu} h \right)\right],
\end{equation}
where $\xi$ is a generic gauge parameter. When $\xi=1$, the above Lagrangian reduces to the background harmonic gauge-fixing Lagrangian used in \cite{r5}.

The corresponding ghost Lagrangian  may be written in the form 
\begin{equation}\label{eq6}
{\cal L}_{gh} = \sqrt{-\bar g}\; c^{* \mu} \left[ \bar D_{\lambda} \bar D^{\lambda}
\bar{g}_{\mu \nu} - \bar R_{\mu \nu} \right] c^{\nu}.
\end{equation}
We note that the above expressions are invariant under the background field transformation \eqref{eq4}.
The Feynman rules for propagators and interaction vertices are given in Appendix A of  Ref. \cite{ra}. 

In order to extend this theory at finite temperature, we will employ the imaginary time formalism 
introduced by Matsubara and developed by several authors \cite{rb,rc,rd,rd1}. The calculation of an amplitude
in this formulation is rather similar to that at zero temperature. The only difference is that the energy, 
instead of taking continuous values, takes discrete values, that ensures the correct periodic boundary 
conditions for bosonic amplitudes (and anti-periodic for fermionic amplitudes) . For example, in the case of graviton self-energy, one has $p_0=2\pi i n T$, $n=0,\pm 1, \pm 2, \dots$. Consequently, when evaluating Feynman loops, the loop energy variable, rather than being integrated, is summed over all possible discrete values. This sum, to one loop, gives rise to a single Bose-Einstein statistical factor
\be\label{eq7}
N\left(\frac{|\vec p|}{T}\right) = \frac{1}{\exp\left(\frac{|\vec p|}{T}\right) - 1}
\ee
The amplitude naturally separates into a zero-temperature and a temperature dependent part. The thermal 
part can be represented as a forward scattering amplitude, where the internal line is cut open to be on-shell with the corresponding statistical factor \cite{re,rf,rg}. The real-time result  can be obtained by an analytical continuation of the external energy $k_0 \rightarrow (1 + i\epsilon) k_0$. This method is calculationally convenient, as will be illustrated in the next section for the graviton self-energy at finite temperature.

\section{The thermal graviton self-energy}\label{sec3}
\begin{figure}[b!]
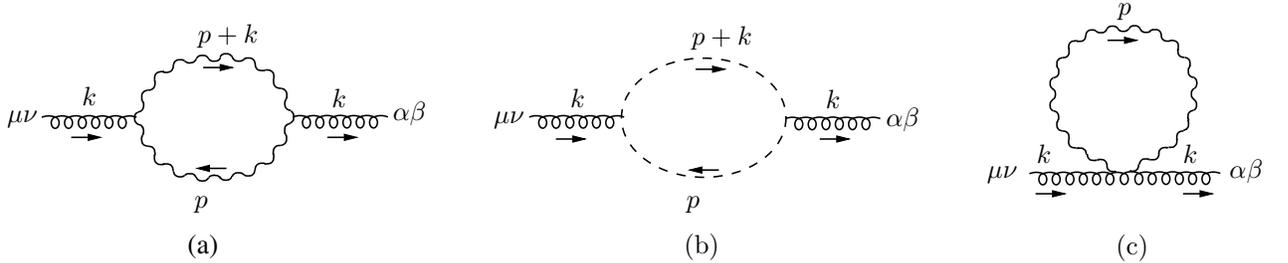

\input{graviton_loopN1.pspdftex}\qquad \qquad
\input{ghost_loop1.pspdftex}\qquad \qquad
\input{tad_pole1.pspdftex}
\caption{One-loop contributions to $\langle \bar h\bar h \rangle$. The curly, wavy and dashed lines are associated with the background fields, the quantum fields and the ghost fields respectively. The arrows indicate the direction of momenta.
}\label{fig1}
\end{figure}

The Feynman diagrams contributing at one loop to the graviton self-energy are indicated in Fig. (\ref{fig1}).
As shown in \cite{ra}, at zero temperature the singular terms for $d = 4 - 2\epsilon$  may be written in the transverse form 
\be\label{eq8}
\Pi^{div}_{\mu\nu ,\,\alpha\beta}(k)  = \frac{\kappa^2}{32\pi^2}
\left[ \frac{1}{\epsilon} - \log(-k^2) \right]
k^4\left\{
4 \, c_1(\xi) \, L_{\mu\nu} L_{\alpha\beta} + c_2(\xi) \left[
L_{\mu\nu} L_{\alpha\beta} + \frac{1}{2} \left(L_{\alpha\mu} L_{\beta\nu} + L_{\alpha\nu} L_{\beta\mu}\right) 
\right]
\right\},
\ee
where $L_{\mu\nu} = \frac{k_\mu k_\nu}{k^2} - \eta_{\mu\nu}$ and $c_1(\xi)$, $c_2(\xi)$ are gauge-dependent constants given by
\be\label{eq9}
c_1(\xi) = \left[\frac{1}{120}+\frac{(\xi-1)^2}{6}\right] ; \;\;\;\; c_2(\xi) = \left[\frac{7}{20}+\frac{\xi(\xi-1)}{3}\right].
\ee
This expression has been obtained by using the fact that in a general background gauge, the result can be expressed in terms of combinations of the following three types of integrals (with $a,b=1,2$)
\begin{equation}\label{eq10}
I_{ab}(k) \equiv 
\int \frac{d^d p}{i(2 \pi)^d} \frac{1}{(p^2)^a [(p+k)^2]^b} =  \frac{(-k^2)^{d/2-a-b}}{((4\pi)^{d/2}}
\frac{\Gamma(a+b-d/2)}{\Gamma(a) \Gamma(b)} \frac{\Gamma(d/2-a) \Gamma(d/2-b)}{\Gamma(d-a-b)} 
\end{equation}
and noticing that their singular contributions may be related in the following way 
\be\label{eq11}
I^{div}_{12}(k)=I^{div}_{21}(k) = \frac{k^2}{2} I^{div}_{22}(k) =  -\frac{1}{k^2} I^{div}_{11}(k) .
\ee
Thus, the singular coefficient in Eq. \eqref{eq8} can be expressed  just in terms of  $I_{11}^{div}$, where
\be\label{eq12}
I^{div}_{11}(k) = \frac{1}{16\pi^2}\left[\frac{1}{\epsilon} - \log(-k^2)\right].
\ee

\begin{figure}[b!]
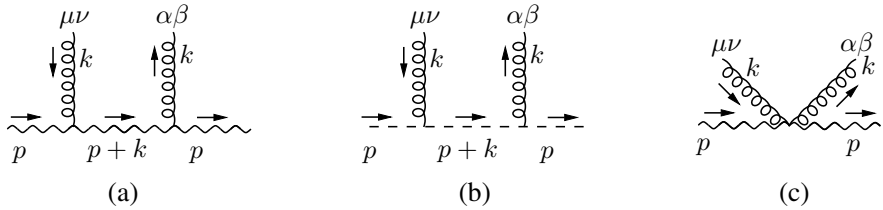

\input{bart1.pspdftex} \qquad \qquad \input{bart2.pspdftex} \qquad \qquad \input{bart3.pspdftex} \qquad \qquad
  \caption{
Forward scattering amplitudes corresponding to Fig. \ref{fig1}. Crossed graphs ($k\rightarrow -k$) are to be understood.
}\label{fig2}
\end{figure}

In order to extend these results at finite temperature, we will express the corresponding contributions
from the diagrams in Fig. \ref{fig1} in terms of the forward scattering amplitudes shown in Fig. \ref{fig2}.
These thermal contributions may be similarly evaluated in terms of the following three types of  temperature dependent integrals ($a,b = 1,2$)
\begin{eqnarray}\label{eq13}
I_{ab}^T(k) &=&-\int \frac{{\rm d}^{3-2\epsilon} p}{(2\pi)^{3-2\epsilon}}  
\left[\frac{1}{({a}-1)!}
\frac{\partial^{{a}-1}}{
\partial p_0^{{a}-1}}\left(\frac{N(p_0/T)}{(p_0+|\vec p|)^{a}}
                   \frac{1}{[(p+k)^2]^{b}}\right) 
\right.
\nonumber \\
&&
\qquad\qquad\qquad
+\left. 
 \frac{1}{({b}-1)!}
\frac{\partial^{{b}-1}}{
\partial p_0^{{b}-1}}\left(\frac{N(p_0/T)}{(p_0+|\vec p|)^{b}}
                           \frac{1}
                           {[(p+k)^2]^{a}}\right)\right]_{p_0 = |\vec p|}
+ (k\rightarrow -k )
\end{eqnarray}
It is possible to evaluate exactly these integrals in terms of logarithmic functions and
of Riemann's  zeta functions \cite{r27a,r27}. Here, our basic interest is to determine the leading thermal
logarithmic contribution which reduces in the zero temperature limit to the $\log(-k^2)$ term in Eq. \eqref{eq8}.
As shown in Appendix \ref{appA}, it turns out that for such a contribution one finds analogous relations to those given in Eq. \eqref{eq11}, namely
\be\label{eq14}
I^{\log T}_{12}(k) = I^{\log T}_{12}(k) = \frac{k^2}{2} I^{\log T}_{22}(k) = -\frac{1}{k^2} I^{\log T}_{11}(k),
\ee
where
\be\label{eq15}
I^{\log T}_{11}(k) = -\frac{2}{(2\pi)^{3-2\epsilon}}\int\frac{d^{3-2\epsilon} p}{2|\vec p|}
\left(\frac{1}{k^2+2 k\cdot p}+\frac{1}{k^2-2 k\cdot p}\right)_{p_0=|\vec p|} N\left(\frac{|\vec p|}{T}\right).
\ee
This shows that in a general background gauge, the leading thermal logarithmic contribution of the graviton self-energy may be expressed just in terms of that arising from $I^{\log T}_{11}$.  
After a straightforward calculation, outlined in the Appendix, we obtain for the corresponding contribution, the result 
\be\label{eq16}
I^{\log T}_{11}(k) = \frac{1}{16\pi^2}\left[
\log(-k^2) - \log(-k^2 - 8\pi i k_0 T + 16\pi^2 T^2).
\right]
\ee
We note that the expression  \eqref{eq16} vanishes in the zero temperature limit, as expected due to the behavior of the statistical factor in Eq. \eqref{eq15}.

Adding the contributions from Eqs. \eqref{eq12} and \eqref{eq16}, one can see that the $\log(-k^2)$ terms cancel out 
since this thermal logarithmic contribution has the same Lorentz form as the one at $T=0$ \cite{r28,r29,r30}. This property, together with the relations \eqref{eq11} and \eqref{eq14}, lead to the conclusion that
the total leading logarithmic contribution coming from the graviton self-energy can be directly obtained by the following extension of the zero-temperature result \eqref{eq8}
\be\label{eq17}
\Pi^{\log T}_{\mu\nu ,\,\alpha\beta}(k)  = -\frac{G}{\pi}
\log(-k^2-8\pi i k_0 T + 16\pi^2 T^2) k^4
\left\{
4 \, c_1(\xi) \, L_{\mu\nu} L_{\alpha\beta} + c_2(\xi) \left[
L_{\mu\nu} L_{\alpha\beta} + \frac{1}{2} \left(L_{\alpha\mu} L_{\beta\nu} + L_{\alpha\nu} L_{\beta\mu}\right) 
\right]
\right\},
\ee
where we have used that $\kappa^2 = 32\pi G$.
This expression has been explicitly verified for the $\log T^2$ contribution which arises at high temperatures.

\section{Quantum corrections to the Newtonian potential}\label{sec4}
\begin{figure}[h!]
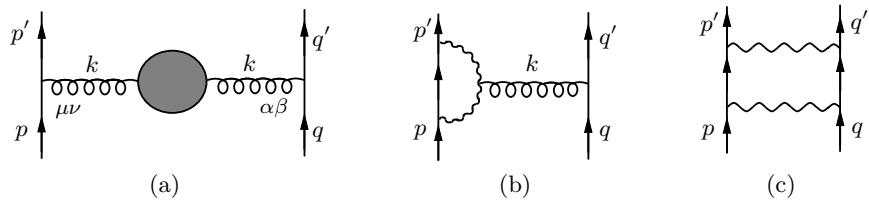

\input potentialA.pspdftex\qquad \qquad
\input potentialB.pspdftex\qquad \qquad
\input potentialC.pspdftex
\caption{Examples of Feynman diagrams which yield corrections to the gravitational potential.}
\label{fig3}
\end{figure}

As an application of the above result, we will evaluate the corrections generated by the thermal graviton self-energy to the classical gravitational potential. To this end, we will proceed similarly to the method used at zero temperature in Ref. \cite{ra}. Thus, we couple the external background field to the energy-momentum tensor $T^{\mu\nu}$ of the matter fields as
\be\label{eq18}
{\cal L}_{I} = -\frac{\kappa}{2} \bar h_{\mu\nu} T^{\mu\nu},
\ee
where we have defined $\bar g_{\mu\nu} = \eta_{\mu\nu}+\kappa\bar h_{\mu\nu}$. For scalar fields described by the Lagrangian
\be\label{eq19}
{\cal L}_{M} =\frac{\sqrt{-g}}{2}\left(g^{\mu\nu}\partial_\mu\phi\partial_\nu\phi - M^2\phi^2\right),
\ee
the energy-momentum tensor is given by
\be\label{eq20}
T_{\mu\nu} = \partial_\mu\phi \, \partial_\nu\phi - \frac{1}{2}\eta_{\mu\nu}(\partial_\lambda\phi \,\partial^\lambda\phi-M^2\phi^2).
\ee
Using this result in Eq. \eqref{eq18}, we obtain in momentum space the graviton-matter coupling 
\be\label{eq21}
V_{\mu\nu}(p,p^\prime)=-\frac{\kappa}{2}\left[p_\mu p^\prime_\nu + p^\prime_\mu p_\nu - \eta_{\mu\nu} (p\cdot p^\prime-M^2)\right].
\ee
We can now calculate the quantum correction coming from the diagram shown in Fig. (\ref{fig3}-a)
This graph yields the contribution (compare with Eq. (3.10) in \cite{ra})
\begin{eqnarray}\label{eq22}
\Delta V^{T}_{self}(k) &=&\frac{V_{\mu\nu}(p,p^\prime)}{2 p_0}
\bar {\cal D}^{\mu\nu, \rho\sigma}\Pi^{\log T}_{\rho\sigma,\,\lambda\delta}
\bar{\cal D}^{\lambda\delta, \alpha\beta}  
\frac{V_{\alpha\beta}(q,q^\prime)}{2 q_0}
\nonumber \\ &=&
\frac{G}{\pi} \ln(-k^2-8\pi i k_0 T+16\pi^2 T^2)
\frac{V_{\mu\nu}(p,p^\prime)}{2 p_0 }
\left[
c_1(\xi) \eta^{\mu\nu} \eta^{\alpha\beta} + c_2(\xi)\frac{\eta^{\alpha\mu}\eta^{\beta\nu}+\eta^{\alpha\nu} \eta^{\beta\mu}}{2}
 \right]\frac{V_{\alpha\beta}(q,q^\prime)}{2 q_0},
\end{eqnarray}
where $\bar{\cal D}^{\mu\nu, \rho\sigma}$ is the background field propagator, $p_0$ and $q_0$ are normalization factors and we have used the transversality of the graviton self-energy.
The thermal part of this propagator involves a $N(|\vec k|/T) \delta(k^2)$ term which yields a vanishing contribution because $\Pi^{\log T}_{\rho\sigma,\,\lambda\delta}$ is proportional to $(k^2)^2$.
%

We will evaluate the above quantity in the case involving two heavy particles with mass $M$, by taking the non-relativistic static limit $p \approx p^\prime\approx(M,0)$ in Eq. \eqref{eq22}. We then get  
\be\label{eq23}
\Delta V^{T}_{self}(k) \approx G^2 M^2\ln({\vec k}^2+16\pi^2 T^2)\left[\frac{43}{15}+\frac{4}{3}(\xi-1)(3\xi-1)\right],
\ee
where we used Eq. \eqref{eq21}, the constants $c_1(\xi)$ and $c_2(\xi)$ given in Eq. \eqref{eq9} and we have set $k_0=0$.
This can be transformed to coordinate space by using the relation \cite{rh}
\be\label{eq24}
\int\frac{d^3 k}{(2\pi)^3} e^{-i\vec k\cdot\vec r} \ln({\vec k}^2+16\pi^2T^2) = -\frac{1}{2\pi}\frac{1}{r^3}
(1+4\pi r T)\exp(-4\pi r T). 
\ee
We thus obtain for the correction generated by the graviton self-energy, the result 
(reinstating factors of $\hbar$ and $c$)
\be\label{eq25}
\Delta V^{T}_{self}(r) = -\left[\frac{43}{30} + \frac{2}{3} (\xi-1)(3\xi-1)\right]
\frac{G M^2}{r}\frac{G\hbar}{\pi c^3 r^2}\left(1+\frac{4\pi rT}{\hbar c}\right)\exp\left(-\frac{4\pi rT}{\hbar c}\right),
\ee
which generalizes the equation (3.13) obtained at zero temperature in Ref. \cite{ra}. 
As explained in this reference, the correction given by the graviton self-energy in the special gauges $\xi=(2\pm\sqrt{13})/3$ matches, at zero temperature, the complete result obtained in the gauge $\xi=1$ in Refs. \cite{r25,r26}. The full correction to the gravitational potential is a physical quantity which is necessarily gauge independent. Moreover, the Fourier transform \eqref{eq24} is, like in the case at $T = 0$, common to all diagrams contributing to the full result. Thus, in these special gauges, the thermal contribution \eqref{eq25} generated by the graviton self-energy yields
\be\label{eq26}
\Delta V^T(r) = -\frac{41}{10} \frac{G M^2}{r}\frac{G \hbar}{\pi c^3 r^2}
\left(1+\frac{4\pi r T}{\hbar c}\right)\exp\left(-\frac{4\pi rT}{\hbar c}\right),
\ee
which gives the complete leading thermal correction to the Newtonian potential.
We note that, in a general background gauge, the physical result obtained in 
Eq. \eqref{eq26} arises only by taking into account the contributions coming from a large number of Feynman diagrams.

\begin{figure}
\includegraphics[scale=0.55]{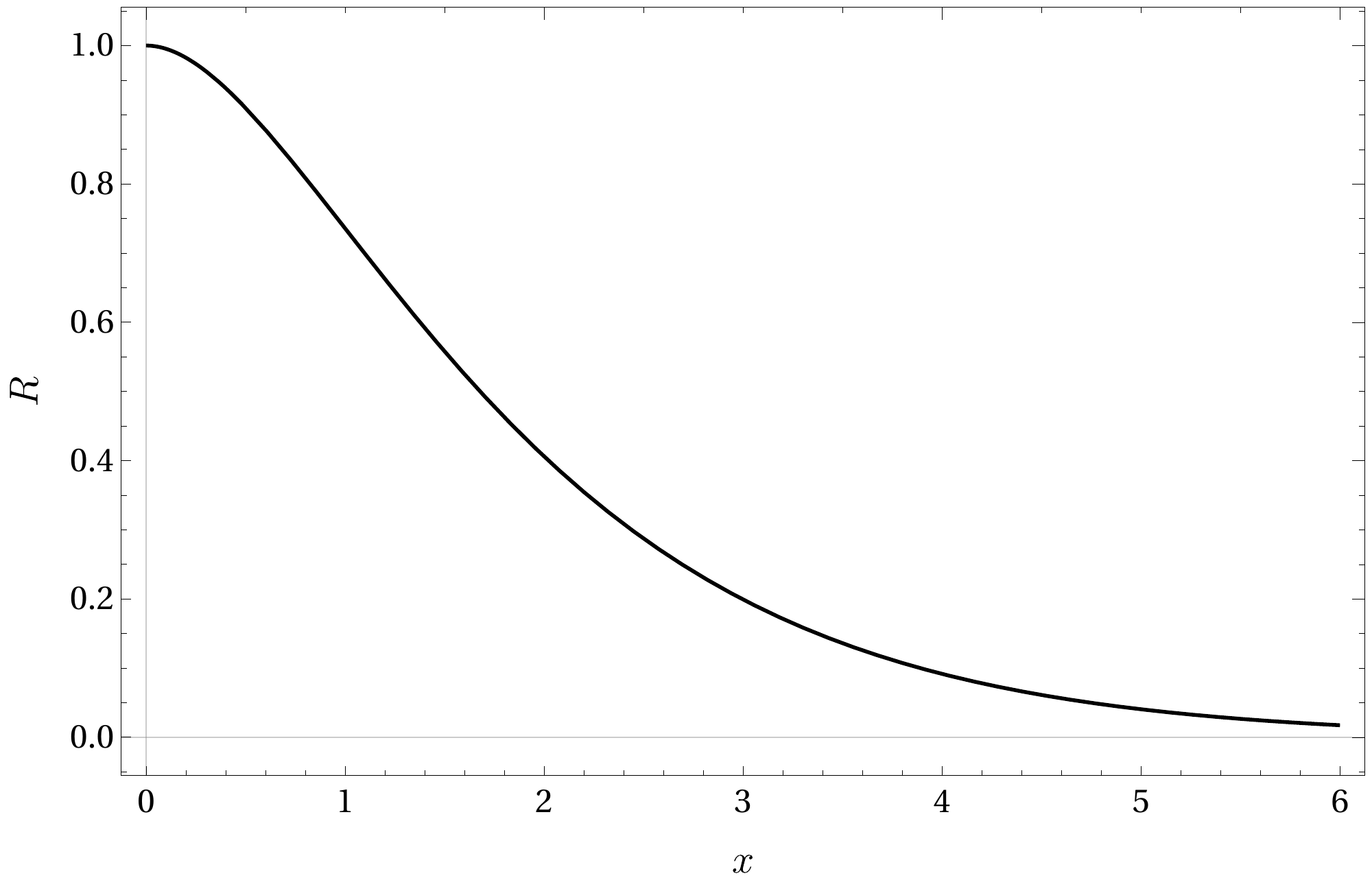}
\caption{The dependence of the ratio $R$ on the parameter $x=4\pi r T/\hbar c$.}\label{fig4}
\end{figure}

A plot of the ratio $R$ between the finite temperature and the zero temperature corrections is shown in Fig. (\ref{fig4}), as a function of the variable $x=4\pi r T/\hbar c$.

\section{Discussion}\label{sec5}
We extended the work done to one-loop order in Ref. \cite{ra} at zero temperature in a
general background gauge, to any finite temperature. We obtained for the leading logarithmic contribution of the thermal graviton self-energy the result given in Eq. \eqref{eq17}, which reduces to that found earlier for the graviton self-energy at $T=0$. The transversality of this term 
is a consequence of the invariance of the theory under background field transformations.
We note that the logarithmic factor in this equation is
gauge-independent, which indicates that its branch cuts may correspond to physical processes that occur at finite temperatures.

We have applied Eq. \eqref{eq17} to evaluate the leading correction to the Newtonian potential generated by the gravitational vacuum polarization at all temperatures. In the special background gauges $\xi=(2\pm\sqrt{13})/3$, we obtained the analytic expression \eqref{eq26} which generalizes the full result previously obtained at $T = 0$.
The quantum factor $G\hbar/c^3 r^2$ is usually very small being about $10^{-38}$ at $r=10^{-15}$ m, but may become appreciable at much shorter distances.
One can see from Eq. \eqref{eq26} 
and from Fig. (\ref{fig4}) that the quantum correction lessens as the temperature increases. This behavior may be understood adapting an argument given by Feynman \cite{ri}. As the temperature rises, the field lines connecting the two particles spread out, because the entropy increases. This broadening of the field configuration reduces the gravitational force between the particles, which leads to a decrease of the magnitude of such corrections.

Thus, in spite of the lack of predictability of quantum gravity at high energy, due to higher-order loops, one can make in this theory calculable physical predictions at low energies.
This
confirms the general effective low energy strategy implemented in the
literature \cite{Burgess:2003jk,r1,r3,r4,Brandt:2023uaw}. (We note parenthetically that there is a proposal for an alternative method of quantizing general relativity, that leads to a renormalizable and unitary theory \cite{r31,r32}. This approach employs a Lagrange multiplier field which restricts the radiative corrections in pure quantum gravity to one-loop order.
Some aspects of thermal quantum gravity have been examined in this context in \cite{Brandt:2021nev}).

\begin{acknowledgments}
We thank \href{https://www.gov.br/cnpq/}{CNPq} (Brazil) for financial support.
\end{acknowledgments}

\appendix
\section{The temperature-dependent integrals $I^T_{ab}(k)$}\label{appA}
We examine here the behavior of the integrals $I^T_{ab}(k)$ defined in Eq. \eqref{eq13}. We begin by considering the integral $I_{11}^T(k)$ given  in  Eq. \eqref{eq15}. In terms of $x=\cos\theta$, where $\theta$
is the angle between $\vec k$ and $\vec p$, we find by setting $\epsilon=0$ and $p\equiv|\vec p|$, that
\be\label{a1}
I_{11}^{T}(k) = 
\frac{k^2}{16\pi^2}\int_{-1}^{1} \frac{dx}{(k_0 - |\vec k| x)^2}\int_0^\infty dp p\frac{N\left(\frac{p}{T}\right)}
{p^2-\frac{1}{4}\left(\frac{k^2}{k_0-|\vec k| x}\right)^2}.
\ee
It is now convenient to make the change of variable
\be\label{a2}
K(x) = \frac{1}{4\pi i}\frac{k^2}{k_0-|\vec k| x};\;\; K_{+}\equiv\frac{k_0+|\vec k|}{4\pi i}
;\;\; K_{-}\equiv\frac{k_0-|\vec k|}{4\pi i}
\ee
so that the above integral may be written in the form  
\be\label{a3}
I_{11}^{T}(k) = 
-\frac{1}{2\pi i|\vec k|}\int_{K{-}}^{K_{+}} dK
\int_0^\infty dp p \frac{N\left(\frac{p}{T}\right)}
{p^2+4\pi^2 K^2}.
\ee
Performing the $p$ integration \cite{rh}, leads to the expression
\be\label{a4}
I_{11}^{T}(k) = 
-\frac{1}{4\pi i|\vec k|}\int_{K{-}}^{K_{+}} dK
\left[\log\left(\frac{K}{T}\right)+\frac{T}{2K} - \psi\left(1+\frac{K}{T}\right)\right]
\ee
where $\psi(x) = d\log \Gamma(x)/dx$ is the  digamma function.
The $K$ integration may be done  by noticing that the $\psi $ function  leads to a surface term.
We thus obtain the  result 
\be\label{a5}
I_{11}^{\log T}(k) = -\frac{1}{16\pi^2}\log\left(\frac{T^2}{-k^2}\right)+\frac{T}{4\pi i |\vec k|}\log\frac{\Gamma(1+K_{+}/T)}{\Gamma(1+K_{-}/T)}.
\ee

We next consider the integral $I^T_{12}(k)$ (see Eq. \eqref{eq13}) 
\be\label{a6}
I_{12}^T(k) = \frac{1}{(2\pi)^{3-2\epsilon}}\int\frac{d^{3-2\epsilon} p}{4 p^2}\left[\left(\frac{N\left(\frac{p}{T}\right)}{p} - \frac{d N\left(\frac{p}{T}\right)}{dp}\right)\frac{1}{k^2+2k\cdot p} +
\frac{4 k_0 N\left(\frac{p}{T}\right)}{ (k^2+2k\cdot p)^2}
  \right] + (k\rightarrow -k)
\ee
It turns out that the leading logarithmic contribution  arises only from the first term
in Eq. \eqref{a6}. This may be evaluated in a similar way to that employed above, which leads to the following equation
\be\label{a7}
I_{12}^{\log T}(k) =
\frac{1}{8\pi i |\vec k|}\int_{K_{-}}^{K_{+}} d K\int_0^\infty d p p^{-1-2\epsilon}\frac{N\left(\frac{p}{T}\right)}{p^2+4\pi^2 K^2}.
\ee
This expression is infrared divergent. Such a divergence arises due to the use of the
integral reduction method, which allows to express the tensor integrals in terms of the 
scalar integrals $I_{ab}(k)$. These divergences cancel in the final result since the graviton self-energy is infrared finite. Thus, we subtract and add to the last term in Eq. \eqref{a7} the part with $p = 0$ in the denominator that leads to an infrared divergence, which will be disregarded due to the above consideration. In the remaining part, we can set $\epsilon=0$, getting  
\be\label{a8}
I_{12}^{\log T}(k) =
-\frac{1}{32\pi^3 i|\vec k|}\int_{K{-}}^{K_{+}} \frac{dK}{K^2}
\int_0^\infty dp p \frac{N\left(\frac{p}{T}\right)}
{p^2+4\pi^2 K^2}.
\ee                                                        
Performing the $p$ integration, we then obtain \cite{rh}
\be\label{a9}
I_{12}^{\log T}(k) =
-\frac{1}{6 4\pi^3 i|\vec k|}\int_{K{-}}^{K_{+}} \frac{dK}{K^2}
\left[\log\left(\frac{K}{T}\right)+\frac{T}{2K} - \psi\left(1+\frac{K}{T}\right)\right]
\ee
We can no longer  integrate the last term in this equation in closed form. But it turns out that the leading logarithmic contribution comes, similarly to the previous case, from the surface term which arises from an integration by parts. Thus, we obtain
\be\label{a10}
I_{12}^{\log T}(k) = \frac{1}{k^2}\left[
\frac{1}{16\pi^2}\log\left(\frac{T^2}{-k^2}\right)-\frac{T}{4\pi i |\vec k|}\log\frac{\Gamma(1+K_{+}/T)}{\Gamma(1+K_{-}/T)}\right].
\ee                                          
We finally consider the integral $I_{22}^T(k)$ in Eq. \eqref{eq13} which leads to 
\be\label{a11}
I_{22}^T(k) = \frac{1}{(2\pi)^{3-2\epsilon}}\int\frac{d^{3-2\epsilon} p}{2 p^2}
\left[\left(\frac{N\left(\frac{p}{T}\right)}{p} - \frac{d N\left(\frac{p}{T}\right)}{dp}\right)\frac{1}{(k^2+2k\cdot p)^2} +
\frac{4 (k_0 + p) N\left(\frac{p}{T}\right)}{ (k^2+2k\cdot p)^3}
  \right] + (k\rightarrow -k)
\ee
Like in the Eq. \eqref{a6}, only the first term turns out to be relevant for our purpose.
This can be computed in a similar way to that used above. After some calculation, we obtain for the leading logarithmic contribution
\be\label{a12}
I_{22}^{\log T}(k) = \frac{2}{k^4}\left[
\frac{1}{16\pi^2}\log\left(\frac{T^2}{-k^2}\right)-\frac{T}{4\pi i |\vec k|}\log\frac{\Gamma(1+K_{+}/T)}{\Gamma(1+K_{-}/T)}\right].
\ee
From the equations \eqref{a5}, \eqref{a10} and \eqref{a12}, one can verify the relation given in
Eq. \eqref{eq14}. Thus, we can write the relevant logarithmic contributions just in terms of those appearing in $I_{11}^{\log T}(k)$.

To proceed, we express the $\log\frac{\Gamma(1+K_{+}/T)}{\Gamma(1+K_{-}/T)}$ term using the series representation \cite{rh}
\be\label{a13}
\log\Gamma(z) = z\log z - z -\frac{1}{2}\log z + \log\sqrt{2\pi}+ \frac{1}{2}\sum_{m=1}^{\infty}\frac{m}{(m+1)(m+2)}\sum_{n=1}^{\infty}\frac{1}{(z+n)^{m+1}}
\ee                           
where $z=1+K_{\pm}/T$. This yields the following logarithmic contributions  
\be\label{a14}
\frac{1}{2}\left(1+\frac{k_0}{2\pi i T}\right)\log\left(\frac{K_{+}+T}{K_{-}+T}\right)+
\frac{|\vec k|}{4\pi i T} \log\left[\left(1+\frac{K_{+}}{T}\right)\left(1+\frac{K_{-}}{T}\right)\right].
\ee
Substituting this expression in Eq. \eqref{a5}, we obtain for the leading logarithmic term
\be\label{a15}
I_{11}^{\log T}(k) \approx -\frac{1}{16\pi^2}\log\frac{-k^2-8\pi i k_0 T+16\pi^2 T^2}{-k^2}.
\ee
We note that this expression vanishes in the zero temperature limit, as expected for the purely thermal contributions due to the statistical  factor \eqref{eq7}. This term yields the contribution shown in
Eq. \eqref{eq16}. After the cancellation of the $\log(-k^2)$ with that present in Eq. \eqref{eq12}, the remaining $\log(-k^2-8\pi i k_0 T+16\pi^2 T^2)$ term can become very large for very small values of $k^2$ and $T^2$. In the static limit, $k_o=0$, such a contribution would dominate over the other contributions arising from Eqs. \eqref{a13} and \eqref{a14}.


\newpage

\begin{thebibliography}{10}

\bibitem{r5}
\href{http://www.numdam.org/item/AIHPA_1974__20_1_69_0.pdf}
{G. 't~Hooft and M.~J.~G. Veltman, Annales Poincare Phys. Theor. {\bf A20},  69 (1974).}


\bibitem{r7}
\href{https://doi.org/10.1016/0550-3213(78)90055-X}
{R.~E.~Kallosh, O.~V.~Tarasov and I.~V.~Tyutin,
Nucl. Phys. B \textbf{137}, 145-163 (1978).}

\bibitem{r8}
\href{https://doi.org/doi:10.1016/0550-3213(85)90243-3}
{D.~M.~Capper, J.~J.~Dulwich and M.~Ramon Medrano,
Nucl. Phys. B \textbf{254}, 737-746 (1985).}

\bibitem{r11}
\href{https://doi.org/doi:10.1016/0550-3213(86)90193-8}
{M.~H.~Goroff and A.~Sagnotti,
Nucl. Phys. B \textbf{266}, 709-736 (1986).}

\bibitem{r12}
\href{https://doi.org/doi:10.1016/0550-3213(92)90011-Y}
{A.~E.~M.~van de Ven,
Nucl. Phys. B \textbf{378}, 309-366 (1992).}


\bibitem{r14}
\href{https://arxiv.org/abs/1701.02422}{
Z.~Bern, H.~H.~Chi, L.~Dixon and A.~Edison,
Phys. Rev. D \textbf{95}, no.4, 046013 (2017).}

\bibitem{Burgess:2003jk}
\href{https://arxiv.org/abs/gr-qc/0311082}
{C.~P.~Burgess,
Living Rev. Rel. \textbf{7}, 5-56 (2004).
}

\bibitem{r1}
\href{https://doi.org/10.48550/arXiv.1507.08194}
{S.~Carlip, D.~W.~Chiou, W.~T.~Ni and R.~Woodard,
Int. J. Mod. Phys. D \textbf{24}, no.11, 1530028 (2015).}


\bibitem{r3}
\href{https://doi.org/10.48550/arXiv.gr-qc/9512024}
{J.~F.~Donoghue,
[arXiv:gr-qc/9512024 [gr-qc]]};
%
\href{https://doi.org/doi:10.4249/scholarpedia.32997}
{Scholarpedia \textbf{12}, no.4, 32997 (2017).}

\bibitem{r4}
\href{https://doi.org/10.48550/arXiv.1702.00319}
{J.~F.~Donoghue, M.~M.~Ivanov and A.~Shkerin,
[arXiv:1702.00319 [hep-th]].}

\bibitem{r25}
\href{https://doi.org/10.48550/arXiv.hep-th/0211072}
{N.~E.~J.~Bjerrum-Bohr, J.~F.~Donoghue and B.~R.~Holstein,
Phys. Rev. D \textbf{67}, 084033 (2003).}

\bibitem{r26}
\href{https://doi.org/10.48550/arXiv.gr-qc/0207118}
{I.~B.~Khriplovich and G.~G.~Kirilin,
J. Exp. Theor. Phys. \textbf{95}, no.6, 981-986 (2002).}

\bibitem{r13New}
\href{https://doi.org/10.1140/epjc/s10052-022-10077-7}
{T. de Paula Netto, L. Modesto, I. L. Shapiro,
Eur. Phys. J. C \textbf{82}, 160 (2022). }


\bibitem{r17}
\href{https://doi.org/doi:10.1103/PhysRevD.12.482}
{H.~Kluberg-Stern and J.~B.~Zuber,
Phys. Rev. D \textbf{12}, 482-488 (1975).}


\bibitem{r19}
\href{https://doi.org/doi:10.1016/0550-3213(81)90371-0}
{L.~F.~Abbott,
Nucl. Phys. B \textbf{185}, 189-203 (1981).}

\bibitem{r20}
\href{https://doi.org/10.48550/arXiv.1705.03480}
{A.~O.~Barvinsky, D.~Blas, M.~Herrero-Valea, S.~M.~Sibiryakov and C.~F.~Steinwachs,
JHEP \textbf{07}, 035 (2018).}

\bibitem{r21}
\href{https://doi.org/10.48550/arXiv.1801.01098}
{J.~Frenkel and J.~C.~Taylor,
Annals Phys. \textbf{389}, 234-238 (2018).}

\bibitem{r22}
\href{https://doi.org/10.48550/arXiv.1905.08296}
{P.~M.~Lavrov, E.~A.~dos Reis, T.~de Paula Netto and I.~L.~Shapiro,
Eur. Phys. J. C \textbf{79}, no.8, 661 (2019).}

\bibitem{r23}
\href{https://doi.org/10.48550/arXiv.1810.10671}
{F.~T.~Brandt, J.~Frenkel and D.~G.~C.~McKeon,
Phys. Rev. D \textbf{99}, no.2, 025003 (2019).}

\bibitem{ra}
\href{https://arxiv.org/abs/2208.13004}
{F.~T.~Brandt, J.~Frenkel and D.~G.~C.~McKeon,
Phys. Rev. D \textbf{106}, no.6, 065010 (2022).}

\bibitem{ra1}
\href{https://inspirehep.net/literature/322527}
{A.~Rebhan,
CERN-TH-6316-91;
}
\href{https://inspirehep.net/literature/317831}
{
Nucl. Phys. B \textbf{368}, 479-508 (1992).}

\bibitem{park1}
\href{https://doi.org/10.3390/particles4040035}
{I.~Y.~Park,
Particles \textbf{4}, no.4, 468-488 (2021).}

  
\bibitem{r28}
\href{https://arxiv.org/abs/hep-th/9803155}
{F.~T.~Brandt and J.~Frenkel,
Phys. Rev. D \textbf{58}, 085012 (1998).}

\bibitem{r29}
\href{https://arxiv.org/abs/0901.3458}
{F.~T.~Brandt, J.~Frenkel and J.~C.~Taylor,
Nucl. Phys. B \textbf{814}, 366-369 (2009).}


\bibitem{r30}
\href{https://arxiv.org/abs/hep-th/9407051v1}
{F.~T.~Brandt and J.~Frenkel,
Phys. Rev. Lett. \textbf{74}, 1705-1707 (1995).}



\bibitem{rb}
\href{https://doi.org/10.1017/CBO9780511535130}
{J.~I.~Kapusta and C.~Gale,
``Finite-temperature field theory: Principles and applications,''
Cambridge University Press, 2011.}


\bibitem{rc}
\href{https://doi.org/10.1017/CBO9780511721700}
{M.~L.~Bellac,
``Thermal Field Theory,''
Cambridge University Press, 2011.}

\bibitem{rd}
\href{https://inspirehep.net/literature/457892}
{A.~K.~Das,
``Finite Temperature Field Theory,''
World Scientific, 1997.}

\bibitem{rd1}
\href{https://arxiv.org/abs/1701.01554}
{M.~Laine and A. Vuorinen,
``Basics of Thermal Field Theory: A Tutorial on Perturbative Computations,''
Springer, 1st ed. 2016.}


\bibitem{re}
\href{https://doi.org/10.1016/0550-3213(92)90480-Y}
{J.~Frenkel and J.~C.~Taylor,
Nucl. Phys. B \textbf{374}, 156-168 (1992).}

\bibitem{rf}
\href{https://doi.org/10.1103/PhysRevD.55.7808}
{F.~T.~Brandt and J.~Frenkel,
Phys. Rev. D \textbf{55}, 7808-7814 (1997).}

\bibitem{rg}
\href{https://doi.org/10.48550/arXiv.2110.07694}
{F.~T.~Brandt, J.~Frenkel, S.~Martins-Filho, D.~G.~C.~McKeon and G.~S.~S.~Sakoda,
Phys. Rev. D \textbf{104}, no.10, 105007 (2021).}

\bibitem{r27a}
\href{https://journals.aps.org/prd/abstract/10.1103/PhysRevD.45.2081}
{A.~P.~de Almeida, J.~Frenkel and J.~C.~Taylor,
Phys. Rev. D \textbf{45}, 2081 (1992)}

\bibitem{r27}
\href{https://arxiv.org/abs/hep-th/9906128}
{F.~T.~Brandt and J.~Frenkel,
Phys. Rev. D \textbf{60}, 107701 (1999).}

\bibitem{rh}
\href{https://archive.org/details/GradshteinI.S.RyzhikI.M.TablesOfIntegralsSeriesAndProducts/page/n9/mode/2up}
{I. S. Gradshteyn and M Ryzhik, ``Tables of Integral Series and Products'', 1980.}

\bibitem{ri}
\href{https://inspirehep.net/literature/168269}
{R.~P.~Feynman,
``A qualitative discussion of quantum chromodynamics in (2+1)-dimensions'', Lisbon, July 9-15, 1981
[PRINT-81-0830 (CAL-TECH)].}

\bibitem{Brandt:2023uaw}
F.~T.~Brandt, J.~Frenkel, D.~G.~C.~McKeon and G.~S.~S.~Sakoda,
Phys. Rev. D \textbf{107}, no.6, 065008 (2023).

\bibitem{r31}
\href{https://arxiv.org/abs/2009.09553}
{F.~T.~Brandt, J.~Frenkel, S.~Martins-Filho and D.~G.~C.~McKeon, 
Annals Phys. \textbf{427}, 168426 (2021)};
\\
\href{https://arxiv.org/abs/2102.02854}
{Annals Phys. \textbf{434}, 168659 (2021)}.

\bibitem{r32}
\href{https://doi.org/10.1016/j.aop.2023.169323}
{F.~T.~Brandt and S.~Martins-Filho, 
Annals Phys. \textbf{453}, 169323 (2023)}.

\bibitem{Brandt:2021nev}
\href{https://arxiv.org/abs/2105.00318}
{F.~T.~Brandt, J.~Frenkel, S.~Martins-Filho, D.~G.~C.~McKeon and G.~S.~S.~Sakoda,
Can. J. Phys. \textbf{100}, no.3, 139-144 (2022).}

\end{thebibliography}
\end{document}